\documentclass[11pt]{article}

\input epsf.sty
\usepackage{color} 
\usepackage[utf8]{inputenc} 

\usepackage{feynmf}
\usepackage{graphics,amsmath,amssymb}
\def\lromn#1{\uppercase\expandafter{\romannumeral#1}}

\begin{document}

\begin{center}
\begin{Large}
\textbf{
B $-$ L genesis by sliding inflaton
}
\end{Large}

\vspace{1cm}

\vspace{1cm}

M. Yoshimura

Research Institute for Interdisciplinary Science,
Okayama University \\
Tsushima-naka 3-1-1 Kita-ku Okayama
700-8530 Japan

\vspace{5cm}

{\bf ABSTRACT}

\end{center}

\vspace{1cm}

We propose a new mechanism of lepton (L) number asymmetry generation,
hence offer an explanation of matter-antimatter imbalance when
a significant amount of baryon number is later
transformed from this L-number by known electroweak sphaleron mediated process.
The basic theoretical framework is a
recently proposed  multiple scalar-tensor gravity that dynamically solves
the cosmological constant problem.
The L-asymmetry generation in one of two proposed scenarios
 is triggered by dynamical relaxation of scalar inflaton field towards 
the zero cosmological constant.
CPT violation (C= charge conjugation, P = parity operation,
T= time reversal) in the presence of a chemical potential 
gives the necessary time arrow, and 
 lepton number violating scattering in cosmic thermal medium 
generates a net cosmological  L-number via resonance formation.
Another scenario is L-asymmetry generation from
evaporating primordial black holes.
These proposed mechanisms do not require CP violating phases
in physics beyond the standard model:
the new  required physics
is existence of heavy Majorana leptons of masses $ 10^{15} \sim 10^{17}$ GeV
 that realizes the seesaw mechanism.
We identify the cosmological epoch of lepto-genesis in two scenarios,
 which may  give the right amount of observed baryon to entropy ratio.
It might even be possible to experimentally determine  microscopic physics parameter, 
masses of three heavy Majorana leptons 
by observing astrophysical footprints of primordial black hole evaporation
at specified hole masses.

\vspace{3cm}

Keywords
\hspace{0.5cm} 
Spontaneous baryo-genesis,
CPT violation in the universe,
Inflaton,
Resonant lepto-genesis,
Heavy Majorana leptons,
Lepton number chemical potential,
Primordial black hole evaporation,
Multiple scalar-tensor gravity,
Cosmological constant problem

\newpage

\section
 {\bf Introduction}

Familiar baryo-genesis \cite{delayed decay},
or lepto-genesis \cite{original lepto-genesis}, scenarios are based 
on what may be called the delayed decay mechanism \cite{delayed decay 2}.
This mechanism uses short time span of the cosmic expansion 
during which heavy particles such as baryon number violating
gauge and Higgs bosons, or right-handed Majorana leptons, decay
without appreciable counter-balance due to inverse decays.
The scenario typically requires the inequality relation between
the thermally averaged decay rate $\Gamma$ and the Hubble rate $H$:
$\Gamma > H > \Gamma_{\rm inv} $ in order to block the inverse decay 
(with its thermally averaged rate $\Gamma_{\rm inv} $).
This inequality is usually regarded as the out-of-equilibrium condition.
It was however later pointed out in \cite{thermal baryo-genesis} that
even balanced baryon or lepton number violating processes
such as scattering among much lighter thermalized particles can generate
the asymmetry provided that a finite chemical potential $\mu$
distinguishing particles from anti-particles is present.
The chemical potential is introduced in grand canonical ensemble 
of statistical physics to cope with the case
when the particle number is not exactly conserved.
An advantage of this new scenario is that the asymmetry may
in principle be accumulated during a longer time span.
We shall use in the present work
 this mechanism within the framework of  new cosmological models.

It was recently shown  \cite{cc relaxation}
that multiple scalar-tensor  gravity solves
major cosmological conundrums, at the same time answering  the
cosmological constant problem.
In the proposed model the inflaton field has two stationary points;
one at a spontaneously broken potential minimum around the Planck scale
 and the other at the field infinity
where a properly defined dynamical cosmological constant vanishes.
In the slow-roll to the  potential minimum, inflation is realized
to give a right amount of the observed spectral index.
The field oscillation around this minimum then
gives violent particle production towards thermalized hot big-bang \cite{preheating}.
The inflaton field oscillation simultaneously accompanies zero-point quantum fluctuation
of another inflaton within the multiple scalar-tensor gravity, 
whose positive contribution may cancel the negative mass squared term 
in a wine-bottle type potential of inflaton fields.
Quantum fluctuation ultimately, but gradually, restores the broken symmetry
of inflaton field system, when the inflaton field moves 
towards the field infinity of zero cosmological constant.

The long-time phase of dynamically relaxing inflaton  is  ideal for fostering
generation of the B$-$L asymmetry according to the idea of
spontaneous B$-$L genesis \cite{thermal baryo-genesis}, in particular,
thermal genesis using a finite chemical potential that
distinguishes particles from anti-particles.
Evaporation of primordial black hole with the presence of a chemical potential
is also a promising site of the asymmetry generation.
We shall discuss these two possibilities and compare their results.

A particular mechanism we detail is based on a standard model extension
adding heavy Majorana leptons (three $N_R$'s) responsible for the seesaw mechanism
of small neutrino mass generation \cite{so10 1}.
Lepton number violating scattering occurs in this model efficiently via
resonance formation of $N_R$'s.
Our scenario does not require CP violating phases in physics beyond the standard model.
CPT violation is provided by a finite chemical potential which we calculate
using standard model parameters in addition to the inflaton sliding velocity.
Necessary physics inputs are all ready for implementation of this lepto-genesis scenario.
We shall be able to locate the energy scale of lepton number violation
by taking the  observed value of baryon to entropy ratio.

The rest of the present paper is organized in the following sections and ends
with a summary.

(2) Cosmological model realizing dynamical relaxation of cosmological
constant; an overview

(3) Generation of chemical potential for particle anti-particle asymmetry

(4) Lepton number violating scattering via resonance

(5) Thermal generation and primordial black hole evaporation

(6) Appendix: Thermal average of scattering rate including
the region of resonance formation

We use the natural unit of $\hbar = c  = 1$ and the Boltzmann constant $ k_B = 1$
throughout the present work  unless otherwise stated.

\section
{\bf Cosmological model realizing dynamical relaxation of cosmological
constant; an overview}

Since the preceding work \cite{cc relaxation} which we base on
is not well publicized, we shall recapitulate main results
relevant to our subsequent discussion of B$-$L genesis in the present work.
Derivation and detailed concepts that led to the preceding work
are relegated to the original paper.

Towards construction of a unified cosmology the present author 
is strongly motivated by a belief that
slow-roll inflation, late time acceleration (dark energy), and hopefully
dark matter should be solved along with the cosmological constant problem
\cite{cc problem rev}.
Otherwise, one is not sure of at which  cosmic epoch
the cosmological constant should be fine-tuned; either the Planck epoch, 
or electroweak epoch,
or QCD epoch, or some other epochs ?
It seems that the phenomenology of cosmology requires at least
two nearly non-varying cosmological constants, one of the Planck order at inflation and
the other of $({\rm meV})^2$ order at present:
are they related or independent ?
There are two important concepts to realize the ultimate  goal in our view:
(1) dynamically varying cosmological constant, (2) restoration of spontaneously broken
symmetry.
We shall spell out these in some detail, adding two comments.

(1) An arbitrary, both in sign and magnitude,
 cosmological constant is introduced in gravity theories
of multiple scalar $\chi$ fields different from general relativity,
in which the Einstein-Hilbert action is changed to allow a lagrantian density
of the form, $- F(\chi) G_N R$, with $R$ the Ricci scalar, $G_N$ the
gravitational constant \cite{jbd}.
The metric frame that has this form is called the Jordan frame.
A positive definite function
$F(\chi)$ is called the conformal function, which may be expanded into a polynomial
of powers, $(\chi/M_{\rm P})^2\,, M_{\rm P}=1/\sqrt{16\pi G_N} \sim 1.7 \times 10^{18}$
GeV.
The inflaton field $\chi$ must  have at least four real components. 
We assume both of conformal function and potential function $V(\chi)$
to possess a global O(4) symmetry.

Physical consequences of cosmology are best worked out
in the Einstein metric frame that can be transformed from the
Jordan frame by a Weyl rescaling, by which $F(\chi)$ factor is eliminated to unity
in the Einstein frame, hence the non-varying gravitational constant emerges in this frame.
An arbitrary cosmological constant $\Lambda$ introduced in the Jordan frame
then becomes a dynamical variable of the form, $\Lambda/F^2(\chi)$,
in the Einstein frame.
This way  the dynamical cosmological 
constant vanishes at the field infinity $\chi = \infty$.
The standard model, or a grand unified model, is
introduced as well in the Jordan frame, resulting in multiplication of some powers
of $1/F(\chi)$ in the Einstein frame after the Weyl scaling.
With time evolution of the inflaton $\chi$ fields, particle physics
parameters  apparently vary, which must be treated with great care,
since the unit change of length and time is also present. 

(2) Important epochs in cosmology can
be regulated by a potential $V(\chi)$ for the infialton field.
The potential is assumed to have a wine-bottle type shape such that
a spontaneous breaking occurs at a field value $ \chi_m$.
There are two corrections to this tree-level potential $V$:
one is the Nambu-Goldstone (NG) kinetic repulsion \cite{ng repulsion},
and the other is quantum fluctuation.
NG kinetic contribution is incorporated as an effective
potential term of centrifugal repulsion equal to $ {\rm constant}^2/a^6(t) \chi^2$,
with $a(t)$ the cosmic scale factor, giving the shifting minimum position 
according to the cosmic expansion.
The first important epoch occurs when the inflaton, starting from
a field point $\chi > \chi_m$,  settles down around
the potential minimum at $\chi_m$.
Due to the inflaton coupling to Higgs field 
$\lambda_H (|H|^2 - v^2)^2/(4F^2(\chi)\,) \,, \lambda_H > 0$ (with $H$ the Higgs doublet,
$v$ vacuum expectation value, and $\lambda_H $ dimensionless coupling)
and other particle fields, we expect that a hot big-bang is realized
using the mechanism of parametric amplification, as outlined in \cite{preheating}.

Effect of growing, though temporarily,
 quantum fluctuation, necessarily accompanying particle production,
may drastically change the nature of O(4) symmetry among four real inflaton fields.
The original negative mass squared term $\propto - \chi_m^2 \chi^2$ in $V(\chi)$
receives a positive-definite correction $ \propto \langle \phi^2 \rangle$
where $\phi$ is an orthogonal component to the dominant $\chi$ inflaton field.
This gives rise to O(4) symmetry restoration, lowering a wall height right to $\chi_m$
and finally making the field infinity the global potential minimum.
The inflaton $\chi$ field then starts to roll down to the potential bottom
at which dynamical cosmological constant vanishes.
It is important that the field zero $\chi=0$ can never
be reached due to that the NG repulsion makes the field origin an infinitely high
wall.

(3) Caveat.
The importance of the metric frame difference in physical
interpretation seems to have been frequently
overlooked in the past literature.
Many popular dark energy models introduce non-minimal kinetic
scalar terms of the form $G(\chi) (\partial \chi)^2$ in the Einstein frame.
These models may be regarded to belong to the same class
of models as the model of \cite{cc relaxation},
when one rescales back to the Jordan frame according to Weyl.
But quite often it is not mentioned in the literature in which frame the
standard particle physics lagrangian and a cosmological constant
are introduced.
The cosmological constant problem becomes a dynamical issue,
only when these are introduced in the Jordan frame, and not in
the Einstein frame.
The Jordan frame seems  more fundamental as
a starting point if four dimensional field theory descends from
higher dimensional Kaluza-Klein unification or superstring theories.

(4) Example of compared metric frame difference.
The metric frame differences have been
analyzed in great detail for simple models in \cite{metric frame comparison}.
In the Jordan-Brans-Dicke theory \cite{jbd}
added by a cosmological constant $\Lambda$,
attractor solutions that give cosmic scale factor $ \propto t^{1/2}$ 
and dynamical cosmological constant $\propto \Lambda/t^2 $ are found in
the Einstein frame.
This model is equivalent to a quintessence model 
\cite{quintessence 1}, \cite{quintessence 2} of potential $\propto \chi^{-2}$
in the Einstein frame, or $F(\chi) \propto \chi^2$ in the Jordan frame.
Surprisingly, this solution corresponds to attractor of the static Minkowski spacetime
in the Jordan frame.
All attractor solutions in the Einstein frame are shown \cite{metric frame comparison}
 to exhibit the dynamical relaxation of cosmological constant.

\vspace{0.5cm}
We need to give some more details of our model for subsequent discussion
of lepto-genesis.
The potential and conformal functions are taken 
to be both quartic in the Jordan frame, 
\begin{eqnarray}
&&
F(\chi) =  1 + \xi_2 (\frac{\chi}{M_{\rm P}})^2 + \xi_4 (\frac{\chi}{M_{\rm P}})^4
\,, \hspace{0.5cm}
\xi_i > 0
\,,
\label {conformal f}
\\ &&
V^{(J)}(\chi) = 
\frac{g}{4} (\chi^2  - \frac{m_{\chi}^2}{g} )^2 + 2 M_{\rm P}^2 \Lambda
+({\rm NG\; centrifugal \; repulsion\; terms})
\,, \hspace{0.5cm}
g > 0
\,.
\label {potential in j-frame}
\end{eqnarray}
This is the simplest scheme for successful realization of ideas stated above.
The basic inflaton field equation (when
a single-field description is applicable) is given by
\begin{eqnarray}
&&
\ddot{\chi} +  3 \frac{\dot{a}}{a} \dot{\chi}
-  \frac{\partial_{\chi} F}{F } \dot{\chi}^2 
= - \partial_{\chi} V_{\rm eff}^{(E)}(\chi)
\,,
\label {conformal chi-eq}
\end{eqnarray}
suppressing contributions from spatially inhomogeneous inflaton fields.
We refer to the original paper \cite{cc relaxation}
for the precise definition of how
the effective potential $ V_{\rm eff}^{(E)}(\chi)$ is related to
$V^{(J)}(\chi)$ and $F(\chi)$, but its
approximate form relevant to our subsequent discussion is given shortly.

The slow-role phase of inflation
occurs in the field region that general relativity approximately
holds, namely at $F(\chi) = 5$.
General relativistic analysis of slow-roll condition
\cite{cosmology textbook} and calculation of
the spectral index along with the tensor to scalar mode ratio
\cite{spectral index etc}
can be applied to constrain the potential derivative and curvature.
We confirmed \cite{cc relaxation} that the slow-roll inflation is realized consistently
with observations \cite{bicep/keck}.

To the rest of discussion on  B$-$L genesis,
it is important to give result of time evolution of inflaton field
in radiation dominant (RD) epoch.
During RD epoch the inflaton field essentially follows a single-field equation,
 and  the effective potential needed for a large $\chi$ field is
\begin{eqnarray}
&&
V_{\rm eff}^{(E)}(\chi) \approx \frac{M_{\rm P}^4}{5}
\left( {\rm constant}  - 
\frac{g}{\xi_4} \ln \frac{\chi}{\chi_*}
\right)
\,, 
\label {effective potential at infty}
\end{eqnarray}
where $\chi_*$ is a starting point of field roll-down to infinity after
the symmetry restoration, and its precise value is irrelevant to our discussion.

The field equation in RD epoch is approximately 
\begin{eqnarray}
&&
\ddot{\chi} +   \frac{\dot{3}}{2t} \dot{\chi}
=  \frac{ g}{5 \xi_4} \frac{M_{\rm P}^4}{\chi}
\,.
\label {conformal chi-eq: asymptotic}
\end{eqnarray}
One arrives at the following solution, supported by some limited
numerical simulations,
\begin{eqnarray}
&&
\chi(t) = \sqrt{\frac{ 2g}{15 \xi_4 }} M_{\rm P}^2 t
\,.
\label {chi-t}
\end{eqnarray}
Namely, the first term $\ddot{\chi}$ and $\partial_{\chi}F/F$
term are negligible compared to the Hubble term
of right-hand side of eq.(\ref{conformal chi-eq}).
Since the redshift factor is defined by $z+1 = a(t_0)/a(t)$ with $t_0$ 
the present cosmic time, this gives a simple relation,
($H_0 =$ the present Hubble rate),
\begin{eqnarray}
&&
\chi(t) = \chi(t_0) (z+1)^{-2}
\,, \hspace{0.5cm}
\chi(t_0) \approx \sqrt{\frac{1 }{ 30}} \sqrt{\frac{g }{ \xi_4}} \frac{M_{\rm P}^2 }{H_0 }
\sim 0.31 \times 10^{60} \sqrt{\frac{g }{ \xi_4}} M_{\rm P}
\,.
\label{chi variation}
\end{eqnarray}
Thus, in RD epoch the inflaton field $\chi$ increases like $\propto T^{-2}$
with the cosmic temperature, as the cosmic time increases.
At later times after recombination the field increases more slowly,
typically $\propto (t/H_0)^2$,
but this behavior  is irrelevant to our discussion of lepto-genesis.
The $F-$factor in RD epoch at the field infinity is approximately
\begin{eqnarray}
&&
F^{-2} \approx (\xi_4 (\frac{\chi}{M_{\rm P}})^4 )^{ -2 }
\propto (z+1)^{16 }
\,.
\end{eqnarray}

It has been argued that cold dark matter is generated
as spatially inhomogeneous components of inflaton fields generated
from thermal cosmic medium \cite{cc relaxation}.
The ratio of gravitational strength to a clump mass $M$
made of $\chi$ cold dark matter, $G_N M^2$, increases
tremendously, as discussed in \cite{strong gravity},
giving a rational of considering the subsequent scenario of black hole
evaporation for the asymmetry generation.
A typical growth of $G_N M^2$ is given by $\propto (z+1)^{16 }$.
One would expect that these strongly bound systems merge into
heavier black holes, emitting primordial gravitational waves.
On the other hand, if the clump is made of ordinary nucleons,
the ratio grows $\propto (z+1)^{16\epsilon }$
where $\epsilon$ was introduced for consistency with
nucleo-synthesis and requirement of local gauge invariant field theories
\cite{strong gravity}.
Even if there exist plenty of primordial black holes,
it is not clear how much fraction of these is made of $\chi$ matters.
In our black hole evaporation scenario
we shall have to assume that there exist sufficiently large amount of
$\chi$ primordial black holes.

\section
{\bf Generation of chemical potential for particle anti-particle asymmetry}

Inflaton field $\chi$ may couple with the B$-$L number current, to give
an effective lagrangian \cite{thermal baryo-genesis} of the form,
\begin{eqnarray}
&&
{\cal L}_{B-L} = \eta \partial_{\mu} \chi J_{B-L}^{\mu}
= -\eta  \chi \partial_{\mu} J_{B-L}^{\mu} + {\rm total\; derivative}
\,,
\end{eqnarray}
where $J_{B-L}^{\mu}$ is the B$-$L number four-current.
$\eta$ has a negative mass dimension of  $-1$, and is suppressed
by a new physics scale beyond the standard particle physics model.
When the B$-$L number is violated with $\partial_{\mu} J_{B-L}^{\mu} \neq 0$,
this CP-odd operator may give rise to the B$-$L asymmetry.
In a spatially homogeneous FRW (Friedmann-Robertson-Walker) metric,
$ds^2 = dt^2 - a^2(t) (d\vec{x})^2$, 
this effective term is equivalent to
\begin{eqnarray}
&&
{\cal L}_{B-L} = \mu_{B-L} n_{B-L}
\,, \hspace{0.5cm}
\mu_{B-L} = \eta \dot{\chi}
\,.
\end{eqnarray}
The quantity $\mu_{B-L}$ is the chemical potential that
distinguishes particles of positive B$-$L from anti-particles of negative value.
The asymmetric number density, namely difference between particles and
anti-particles, is given by
\begin{eqnarray}
&&
\frac{1}{3} \mu_{B-L} T^2
\,,
\label {asymm n-density}
\end{eqnarray}
for the chemical potential $\mu_{B-L} \ll T$.
In the model of \cite{cc relaxation} the inflaton
velocity $\dot{\chi} $ is maximally large and of the Planck order, $ O(M_{\rm P}^2)$,
as seen from (\ref{chi-t}).
The generated B$-$L number at early epochs of cosmological evolution
 is re-shaffled via sphaleron \cite{sphaleron} mediated process that
later occurs \cite{spahleron process}, \cite{spahleron rate}.

It was pointed out that if the generation of chemical potential
occurs in the phase of  scalar field 
oscillation  around a potential minimum,
the generated asymmetry is wiped out \cite{oscillational cancellation}.
But our proposed mechanism here occurs in the phase of monotonic
inflaton sliding of $\dot{\chi} > 0$, and this criticism does not apply.
The monotonic roll down to the field infinity
is an essential element of resolving the cosmological constant problem
in the scheme of \cite{cc relaxation}.

The chemical potential in statistical physics is defined in the hamiltonian
formalism.
In \cite{chemical potential} it was shown that the proper chemical potential
thus defined may differ from $\dot{\theta}$ used here and
in the literature \cite{thermal baryo-genesis}.
But the use of $\dot{\theta}$ for discussion of B$-$L genesis remains
valid, hence we shall keep this terminology for a mere convenience.

This mechanism of B$-$L genesis  has features distinct
from the familiar delayed decay mechanism \cite{delayed decay}.
In general this makes the generated asymmetry larger because
one does not need CP violating phases that necessarily
arises in higher orders of perturbation \cite{delayed decay 2}.
The need for time arrow is provided by 
a non-vanishing time derivative of inflaton sliding velocity
$\mu_{B-L} \propto \dot{\chi} \neq 0$.
The time span or temperature span is given by the temperature range
of inequality  $T > T_D$ when this mechanism works, and
it is only limited by the decoupling temperature $T_D$.

\vspace{0.5cm}
\hspace*{5cm}
\begin{fmffile}{vertex10}
\begin{fmfchar*}(30,30)
\fmfstraight
\fmfleft{i1}
\fmfright{o1,o3}

\fmf{scalar,label=$\chi$}{i1,v1}
\fmf{photon,label=$H$}{v1,v2}
\fmf{photon,label=$H$}{v1,v3}
\fmf{fermion,label=$l$}{v2,o1}
\fmf{fermion,label=$l$}{o3,v3}
\fmf{fermion,label=$l$}{v3,v2}

\end{fmfchar*}

\end{fmffile}

\begin{figure*}[htbp]
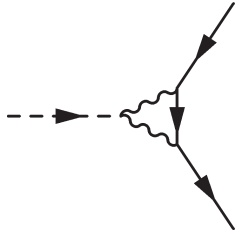

 \begin{center}
 \epsfxsize=0.6\textwidth
   \caption{
Effective $\chi l \bar{l} \; (l= e\,, \mu\,, \tau)$ vertex diagram for chemical potential
$\mu_{B-L}$. Wavy circulating line is neutral Higgs boson $H$. 
}
\label{chi-vertex}
 \end{center} 
\end{figure*}

The chemical potential is calculated from a triangular one-loop diagram,
as shown in Fig(\ref{chi-vertex}),
for the effective vertex of three external fields, $\chi \bar{l} l$
with $l$ one of charged leptons, $e\,, \mu\,, \tau$.
Its probability amplitude has the form in the low energy limit,
\begin{eqnarray}
&&
{\cal A}_ 0 =
 \frac{i }{2 (4\pi)^2} c_{\chi H} y_{Hl}^2
 m_{l} \bar{l} l
\int_0^1 dx \frac{(1-x)^2}{ m_H^2 (1-x) + m_l^2 x}
\,,
\end{eqnarray}
where $m_H$ is the Higgs boson mass, $m_l$ is a lepton mass,
$y_{Hl} = m_l/v\,, v= 254$GeV$\times \sqrt{2}$ is the Yukawa coupling.
The parameter $c_{\chi H}$ arises from inflaton coupling to Higgs boson
of a vertex form $\chi H H$,
\cite{cc relaxation} 
\begin{eqnarray}
&&
V_{\chi H}^{(E)}  = \frac{\lambda_H}{4 F^2(\chi)} (|H|^2 - v^2)^2
\,, \hspace{0.5cm}
c_{\chi H} =   8  \lambda_H \frac{v^2}{\chi} (\xi_4
\frac{\chi^4}{M_{\rm P}^4} )^{-2}
\,.
\label {inflaton coupling to h}
\end{eqnarray}
The quantity $\chi/M_{\rm P}$ evolves with cosmic time $t$ as given
by (\ref{chi-t}); $\chi/M_{\rm P} = O(M_{\rm P} t)$ in RD epoch.
Parameters $g\,, \xi_4$ are dimensionless coupling constants of order unity
given in (\ref{conformal f}) and (\ref{potential in j-frame}).
The time evolution of inflaton field is given by the redshift dependence 
$\propto (z+1)^{-2}$ of sliding inflaton $\chi(t)$.
The redshift is defined by the ratio of scale factor to its present value,
$z+1 = a(t_0)/a(t)$ and is proportional to the cosmic temperature
ratio $T/T_0$.
The integral is simplified for a mass relation $m_H \gg m_l$:
\begin{eqnarray}
&&
\int_0^1 dx \frac{(1-x)^2}{ m_H^2 (1-x) + m_l^2 x} \sim
\frac{1}{ 2 m_H^2} 
\,.
\end{eqnarray}
The  probability amplitude is then given by
\begin{eqnarray}
&&
{\cal A}_0 \sim 
 \frac{i }{4 (4\pi)^2} \frac{c_{\chi H}}{m_H^2} \sum y_{Hl}^2
 m_{l} \bar{l} l
\,.
\end{eqnarray}
It is interesting that this formula depends on known parameters of
electroweak theory except $c_{\chi H}$ of (\ref{inflaton coupling to h}).
The dominant contribution is from a largest $y_{Hl}$,
hence from the coupling to the heaviest charged lepton, $\tau$ pair. 
The chemical potential is thus
\begin{eqnarray}
&&
\mu_L = 
 \frac{c_{\chi H} }{4 (4\pi)^2}  
\frac{m_{\tau}^2}{m_H^2 v^2} 
\dot{\chi}
\,, \hspace{0.5cm}
\frac{c_{\chi H} \dot{\chi}}{v^2} = 8  \lambda_H (\xi_4)^{-2}
(\frac{\chi}{M_{\rm P}} )^{-9}
\sqrt{\frac{2g }{15 \xi_4 }} M_{\rm P}
\,.
\label {lepton chemical potential}
\end{eqnarray}
We replaced the general notation $\mu_{B-L}$ in previous formulas by
the leptonic chemical potential $\mu_L$ to better fit the present model.
This chemical potential
 is extremely sensitive to the inflaton rolling of $\chi$, 
hence is rapidly decreasing function
of redshift factor $\propto (z+1)^{-18 } \propto T^{18}$.
The high power of this behavior arises from $c_{\chi H}$ of (\ref{inflaton coupling to h}),
and  the inflaton linear velocity $ \dot{\chi}$ in $\mu_L$ 
gives no redshift dependence as clear from (\ref{chi-t}).

\section
{\bf Lepton number violating scattering via resonance}

We shall discuss in the following  section
 two possible scenarios of lepton number generation:
firstly thermal genesis and secondly primordial black hole evaporation.
We need a common source of
lepton number violating process in these two scenarios,
and for this matter we need physics beyond the standard particle physics theory.

We consider SO(10) related models \cite{so10 1}, \cite{so10 2}.
More specifically, it is not even necessary to completely specify
SO(10) models, and any SO(10) model that introduces the Majorana $N_R$
mass term violating the lepton number and Dirac-type mixing
arising from its Higgs coupling to
left-handed neutrino field of the form,
\begin{eqnarray}
&&
\left(
N_R^{T} \,{\cal M}\, i \sigma_2 \sigma \cdot \partial \, N_R + ({\rm h.c.}) \right)
+ \left( N_R^{\dagger} m \nu_L + ({\rm h.c.}) \right)
\,,
\end{eqnarray}
is all we need.
We used two-component spinor fields with $\sigma = (1, \vec{\sigma})$. 
Diagonal elements $M_N$ of the $3\times 3$
 matrix ${\cal M}$ are assumed to be much larger than
the standard model scale of 1 TeV, presumably close to grand unification scale
or the Planck scale $M_{\rm P}$.
For a single flavor case the mass diagonalization of
the neutral lepton sector provides the seesaw mechanism \cite{so10 1},
giving a tiny Majorana mass of order, $m_{\nu} = m^2/M_N$, for ordinary neutrinos.
Heavy Majorana particles stay with their masses of order $M_N$.
We thus assume only a minimum extension of standard model to accommodate
B$-$L violation based on already experimentally established neutrino oscillation.

Actually, existing neutrino oscillation data \cite{pdg 22}
suggest that there may be a hierarchy of three neutrino masses,
the heaviest having a larger mass than of order 50 meV, 
the next heaviest of order 10 meV,
and the lightest mass being unknown.
There is a plenty of room accommodating the situation like this one,
since many parameters, three diagonal elements $M_N$, $m$'s, and
the mixing matrix elements, combine to define ordinary neutrino masses.

\begin{figure*}[htbp]
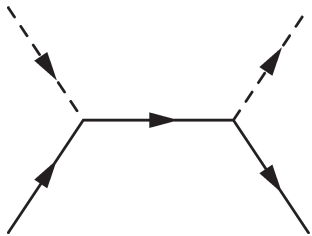

 \begin{center}
 \epsfxsize=0.6\textwidth
   \caption{
Diagram for lepton-number violating scattering $l H^+ \rightarrow \bar{l} H^-$
via intermediate $N_R$.
Another $u-$channel exchange diagram contributing to (\ref{l-violating amp})
 is not depicted for simplicity.
}
\label{l-violating scattering}
 \end{center} 
\end{figure*}

A finite chemical potential supported by lepton number violating process
gives a basis of  asymmetry generation.
As the major process of lepton number violation that occurs till later epochs
one can think of inelastic scattering of charge exchange,
$l + W^{+} (H^+) \rightarrow \bar{l} + W^-(H^-)$,
with $W^{\pm} ( H^{\pm})$ charged weak gauge (Higgs) boson, with
its Feynman diagram shown in Fig(\ref{l-violating scattering}).
The charge exchange scattering amplitude is proportional to
the Majorana mass $M_N$, while the associated elastic scattering amplitude
is proportional to kinetic term. 
Prior to the electroweak phase transition the longitudinal part
of $W^{\pm} $, namely the absorbed part $H^{\pm}$ in the minimum
Higgs doublet model, contributes.
Since the cross sections for lepton number decreasing  $l \rightarrow \bar{l}$
and increasing $\bar{l} \rightarrow l$ processes are equal at the tree diagram level,
the net generated asymmetry is given by the difference of thermal phase space factors
with different signs of the chemical potential, as given by (\ref{asymm n-density}).

The probability amplitude of lepton-number
violating process $l + W^{+} (H^+) \rightarrow \bar{l} + W^-(H^-)$ 
via Majorana $N_R$ exchange is, in the free space,
\begin{eqnarray}
&&
{\cal A}_L = - i y_{HN}^2
 M_N \bar{\psi_v^c} \left( \frac{ 1}{s - M_N^2 } +  \frac{ 1}{u - M_N^2 } \right) \psi_u
\,,
\label {l-violating amp}
\end{eqnarray}
in terms of invariant squared momentum sum, 
$s = (p_l + p_H)^2\,, u = (p_l - p_{\bar{l}})^2$.
$\psi_u$ and $\psi_v$ are four-spinor plane wave functions
of lepton and anti-lepton.
This amplitude is for a single $N_R$, and one must sum over
three species of Majorana leptons.
$y_{HN}$ is the bilinear Higgs coupling to $ \bar{\nu}_L N_R$.
From this probability amplitude one calculates the cross section
in the free space, and takes the thermal average over initial leptons and
Higgs bosons at temperature $T$.

Thermally averaged 
cross sections differ depending on whether the temperature is in the
mass region, $T \approx M_N$, or outside this region.
The thermal average is calculated and numerically illustrated
in Appendix. We shall mention main results here.
The low energy limit $T/M_N \rightarrow 0$ has a finite value 
independent of temperature, of order
$y_{HN}^4 /M_N^2 $ as given by (\ref{low energy sigma}),
while the high energy limit $T/M_N \rightarrow \infty$ 
decreases as a temperature power $\propto 1/T^4 $
as given by (\ref{high energy sigma}).
The most important  is in the resonance energy region
 at $s= M_N^2$.
The resonance temperature is estimated as $T_R \approx M_N/3$,
as given by (\ref{resonance temperature}).
The resonance width is $\Delta T \approx 0.086 \gamma_N$
in terms of natural + thermal width $\gamma_N$;
a quantity not easy to calculate at present.
The cross section in the narrow width limit
behaves as a Lorentzian function almost like a Dirac delta-function,
giving a large resonance cross section $\propto 1/\Delta T$.
The behavior of thermally averaged cross section
is illustrated for large widths (numerically difficult to work out narrow width cases)
in Fig(\ref{l-violating acattering rate}).

\vspace{1cm}
\hspace*{5cm}
\begin{fmffile}{scatt3}
\begin{fmfchar*}(40,30)
\fmfstraight
\fmfleft{i1,i2}
\fmfright{o1,o2}

\fmf{fermion,label=$l$}{i1,v1}
\fmf{scalar,label=$H$}{i2,v1}

\fmf{fermion,label=$N$}{v1,v2}

\fmf{fermion,label=$\bar{l}$}{v2,o1}
\fmf{scalar,label=$H$}{v2,o2}

\end{fmfchar*}
\end{fmffile}

\section
{\bf Thermal generation and primordial black hole evaporation}

\subsection
{\bf Thermally generated lepton asymmetry}

Before we discuss the main issue of this section, we first note that
the inflaton time evolution  in Section \lromn2 is
conveniently written in terms of the Planck energy 
$M_{\rm P} \sim 1.7 \times 10^{27}$ eV,
the present Hubble rate $H_0 \sim 1.6 \times 10^{-33}$ eV,
the present microwave temperature $T_0 \sim 2.35 \times 10^{-4}$ eV,
and the temperature in radiation dominated (RD) epoch $T$;
\begin{eqnarray}
&&
\left( \frac{\chi}{M_{\rm P}} \right)^{-1} = \sqrt{30} \sqrt{\frac{\xi_4}{g}}
\frac{ H_0 T^2}{ M_{\rm P} T_0^2} \sim 9.5 \, \sqrt{\frac{\xi_4}{g}} \left(
\frac{ T}{10^{17} {\rm GeV} } \right)^2
\,.
\label {chi vs temperature}
\end{eqnarray}

First, we discuss the temperature region in which
lepton number violating scattering works.
In cosmic thermal medium L-violating scattering process 
decouples at a temperature $T_D$
derived from the relation $\Gamma_L = 3 H$ 
with $H \propto T^2/M_{\rm P}$ the Hubble rate.
The high power dependence of $\Gamma_L \propto \mu_L \propto (\chi/M_{\rm P})^{-9}$
makes 
the decoupling temperature depend on parameters of the model unusually,
\begin{eqnarray}
&&
T_D = 1.4 \times10^{16} {\rm GeV}\,
\left( \frac{\gamma_N }{ M_N} \right)^{1/16 }
\left(\frac{ \xi_4^2}{ \lambda_H} \sqrt{\frac{\xi_4 }{ g}  }
\right)^{1/16} N_{\rm eff}^{1/32 }
\,.
\label {decoupling temperature}
\end{eqnarray}
We have used the resonance formation cross section $\sigma$
in the rate formula $\Gamma_L = \sigma \mu_L T^2/6$,
calculated in Appendix; eqs.(\ref{l-violating cross section})  and (\ref{resonance integral}).
$\gamma_N$ is the resonance width.
The right-hand side value is insensitive to details of model parameters.
There are three resonances, $\sim M_N^i/3\,, i=1,2,3 $ 
corresponding to three diagonal mass matrix elements.
The decoupling temperature $T_D$ should be set below
the smallest resonance temperature $T_R \approx M_N/3 $
in order to fully exploit large resonance contributions.

We next turn to the amount of generated asymmetry.
The instantaneous L-asymmetry 
generation rate $\Gamma_L$ is given in terms of thermally averaged cross section
$\sigma(T)$,
\begin{eqnarray}
&&
\Gamma_L (T) =  \sigma(T) \frac{\mu_L T^2}{6} 
\,.
\end{eqnarray}
As already stated, this rate may accumulate during a long time span,
and the accumulated asymmetry is given by
\begin{eqnarray}
&&
{\cal Y} = \int_0^{t_D} dt\, \Gamma_L(T) = 2 \int_{T_D}^{\infty} 
dT\, \frac{t(T)}{T} \Gamma_L(T)
= \frac{2}{\pi} \sqrt{\frac{45 }{N_{\rm eff} }} M_{\rm P}
\int_{T_D}^{\infty} dT\, \frac{ \Gamma_L(T)}{T^3}
\,.
\end{eqnarray}
Here $N_{\rm eff}  $ of order 100 is 
the effective massless degrees of freedom contributing
to cosmic energy density.
This formula is valid in RD epoch in which
$t \propto M_{\rm P}/T^2$.

Assuming dominant contribution to arise from the resonance
region $T = T_R \approx M_N$, one may approximate by the integral here,
\begin{eqnarray}
&&
\int_{T_D}^{\infty} dT\, \frac{ \Gamma_L(T)}{T^3} \approx
\frac{1}{6} \sigma_R \, \Delta T \left( \frac{\mu_L}{T} \right)_{T = T_R}
\,.
\end{eqnarray}
From formulas in the preceding sections, we may write
\begin{eqnarray}
&&
\sigma_R \, \Delta T \approx \frac{8.77}{ 8 \pi^4} y_{HN}^4 
\frac{M_N  \Delta T}{\gamma_N T_R^2 } \sim 0.84 \times 10^{-2}\,
\frac{y_{HN}^4}{ M_N}
\,,
\\ &&
 \left( \frac{\mu_L}{T} \right)_{T = T_R} \approx
3.2 \times 10^{-9}\,  \frac{1}{\sqrt{N_{\rm eff} }} \sqrt{\frac{g}{\xi_4}} 
\frac{\lambda_H}{\xi_4^2} \left( \frac{10^{16}{\rm GeV} }{M_N } \right)^2
\,.
\end{eqnarray}
From these we derive
\begin{eqnarray}
&&
{\cal Y} \approx 3.3 \times 10^{-9}\,  \frac{y_{HN}^4}{N_{\rm eff}}
\left( \frac{10^{16} {\rm GeV} }{ M_N} \right)^3
\sqrt{\frac{ g}{\xi_4}} \frac{\lambda_H} {\xi_4^2} 
\,.
\label {b-asymmetry in thermal genesis}
\end{eqnarray}
Due to three resonances this number is essentially tripled.
The lepton/entropy ratio is down by $n_L/s$ with the entropy density
$s= 2 (s_B + 7 s_F/8) T^3/3$.
With the electroweak baryon non-conservation \cite{spahleron process}
 taken into account,
the remaining baryon to entropy ratio is estimated to be $\mu_L/T_D$ times
$O(0.1) \frac{28}{79} $ \cite{b non-conservation at electroweak}.
There is thus a possibility to explain the observed baryon asymmetry.
Our result ${\cal Y}$ weakly depends on the Majorana mass, $\propto M_N^{-3}$.

Resonance formation kinematically requires high ambient temperatures larger than
the mass of heavy Majorana particle; $T > O(M_N)$.
On the other hand, lower reheating temperatures are demanded by
the density perturbation generated at inflation.
The decoupling temperature given by (20)
assuming resonance formation appears a bit large from this point,
although a more precise formula of inflaton time evolution than given by 
(19) is to be used to arrive at a definite conclusion.
In view of this we  explored the possibility of 
low ambient temperatures taking lepton number
violating scattering off the resonance region.
Using low temperature limit of scattering cross section, one derives
decoupling temperature,
\begin{eqnarray}
&&
T_D = 0.95 \times 10^{15} {\rm GeV} \,N_{\rm eff}^{1/36}
y_{HN}^{-2/9} \left( \frac{g^2}{\xi_4 \sqrt{\lambda_H} }\right)^{1/9}
\,.
\end{eqnarray}
This presumably clears the condition of low reheat temperature for the density
perturbation.
The generated baryon asymmetry is calculated as
\begin{eqnarray}
&&
{\cal Y} \approx 3.8 \times 10^{-15} N_{\rm eff}^{-1/2} 
(\frac{10^{16}{\rm GeV} }{M_N })^{2} (\frac{T_D }{ 10^{16}{\rm GeV}})^{18}
y_{HN}^4 \frac{\lambda_H\xi_4^2 }{ g^4}
\,.
\end{eqnarray}
This magnitude is too small compared to the observed value.
An intermediate temperature region between the resonance formation
and the low temperature limit may give a good solution,
but we postpone this study in view of missing improved
treatment of inflaton time evolution.

\subsection
{\bf Asymmetry generated by primordial black hole evaporation}

We turn to the scenario of black hole evaporation.
The idea of using primordial black hole (PBH) evaporation to generate the baryon asymmetry is old \cite{baryo-genesis from bh 1} $\sim $ \cite{bh b-genesis 2}.
We mention a further benefit of our approach over the old scenario:
the inflaton field in the model of \cite{cc relaxation} may gravitationally collapse
to readily form primordial black holes due to  stronger gravity \cite{strong gravity}
than general relativity would predict.
This way  primordial black holes can make up the majority of cold dark
matter, while some fraction may generate the lepton asymmetry.

We first recapitulate main features of black hole evaporation
according to \cite{baryo-genesis from bh 0}, \cite{baryo-genesis from bh 1}, 
\cite{baryo-genesis from bh 2}.
Black holes evaporate thermally with a temperature known as
Gibbons-Hawking temperature related to the hole mass $M_{BH}$,
\begin{eqnarray}
&&
T_{BH} = \frac{16 \pi M_{\rm P}^2}{M_{BH}} \sim 2.65 \times 10^{17}
{\rm GeV} \, \frac{ {\rm mg}}{M_{BH} } 
\,.
\label {gh temperature}
\end{eqnarray}
It is a well-publicized remarkable feature
that smaller black holes evaporate heavier particles
more than lighter ones,
but at a temperature very close to the lifetime it is not entirely clear that
this description remains true due to many fundamental
issues related to information loss paradox.
Due to this Hawking radiation, the hole mass changes with time
according to
\begin{eqnarray}
&&
M_{BH}(t) = M_{BH} ( 1 - \frac{t}{\tau_{BH}})^{1/3}
\,, \hspace{0.5cm}
\tau_{BH} = \frac{20}{\pi N_{\rm eff}} \frac{M_{BH}^3 }{ M_{\rm P}^4 }
\approx  27  \, {\rm y}\, \left( \frac{ M_{BH} } { 10^{12} {\rm gr} 
( N_{\rm eff}/100 )^{1/3}} \right)^3
\,,
\end{eqnarray}
with $M_{BH}$ the initial hole mass at $t=0$.
Although this is not the exponential decay law,
one often calls $\tau_{BH}$ the black hole lifetime.
The mass of PBH's evaporating now is estimated by
equating $\tau_{BH}$ to the present age $\sim 13.5$ Gy, to
give $M_{BH} \sim 8.0 \times 10^{14} {\rm gr}\, (N_{\rm eff}/100)^{1/3}$.
Observation of $\sim $ 1 year gives a history
of PBH evaporation for a hole mass $\sim3 \times 10^{11}$ gr.

There is an important kinetic constraint for successful PBH evaporation
scenario:
the evaporation rate of order $1/\tau_{BH}$ must be larger than
the cosmic expansion rate given by the Hubble rate.
Equivalently this constraint is given by $\tau_{BH} \ll t$ (cosmic time).
This leads to the constraint on cosmic temperature $T$ in RD epoch,
\begin{eqnarray}
&&
\frac{20}{\pi N_{\rm eff}} \frac{M_{BH}^3 }{ M_{\rm P}^4 }
> \sqrt{\frac{45 }{\pi^2  N_{\rm eff} }} \frac{M_{\rm P}}{T^2}
\,.
\end{eqnarray}
This gives a maximum cosmic (not black hole) temperature 
\begin{eqnarray}
&&
T < 3.15 \times 10^{18}{\rm GeV}\,
\left( \frac{M_{\rm P} }{ M_{BH} }\right)^{3/2 } 
\left( \frac{N_{\rm eff}} { 100}\right)^{1/4 }
\,.
\end{eqnarray}
For 1 mg mass of the hole, the right-hand value is 
$1.7 \times 10^{17}{\rm GeV} (N_{\rm eff}/ 100)^{1/4 }$.
It is interesting that this temperature range covers the energy range
of resonance production discussed in the preceding subsection.

The asymmetry instantaneously generated by a single black hole 
is given by ratio of asymmetric number density to entropy density:
\begin{eqnarray}
&&
c \frac{\mu_L}{T_{BH}} =  c\, \mu_L G_N M_{BH}
\,, \hspace{0.5cm}
c \approx \frac{1}{ 2 s_B + 7 s_F/4} = O(0.01)
\,.
\end{eqnarray}
Although the rate of L-number violating scattering $\propto \sigma$
of the preceding section does not enter in this formula, 
its relevance becomes evident when one discusses at which PBH
temperature $T_{BH}$ the asymmetry generation occurs.
The baryon to photon number ratio in the universe is this number times 
the fraction $f$ of PBH abundance.

The instantaneous L-asymmetry generation rate accumulates during the whole
epoch of PBH evaporation, and this gives the accumulated averaged asymmetry,
\begin{eqnarray}
&&
{\cal Y}_{BH} 
= \frac{ 4c}{3} G_N M_{BH} \int_{0}^{ \tau_{BH} } \frac{dt}{\tau_{BH} } \mu_L 
( 1 - \frac{t}{\tau_{BH}})^{1/3}
\,.
\end{eqnarray}
The chemical potential given in (\ref{lepton chemical potential}) 
contains the parameter combination $c_{\chi H} \dot{\chi}$ dependent
on relevant temperature of medium environment.
This can be expressed using (\ref {chi vs temperature}),
\begin{eqnarray}
&&
\hspace*{-0.5cm}
\frac{c_{\chi H} \dot{\chi}}{v^2 M_{\rm P} } = 8\cdot 9.5^9 \,  \lambda_H \xi_4^{-2}
\sqrt{\frac{2g }{15 \xi_4 }} 
  (\frac{\xi_4}{g} )^{9/2} \left(
\frac{ T_{BH} }{10^{17} {\rm GeV} } \right)^{18}
\sim 1.8 \times 10^9 \, \lambda_H \xi_4^{-2} 
(\frac{\xi_4}{g} )^{4} \left(
\frac{ T_{BH} }{10^{17} {\rm GeV} } \right)^{18}
\,,
\\ &&
\mu_L = 5.9 \times 10^{5} M_{\rm P} 
(\frac{\xi_4}{g} )^{4} \left(
\frac{ T_{BH} }{10^{17} {\rm GeV} } \right)^{18}
\sim 1.3 \times 10^{35} \,{\rm GeV}\, (\frac{\xi_4}{g} )^{4} 
\left( \frac{ {\rm mg}}{M_{BH} } \right)^{18}
\,,
\end{eqnarray}
using $(m_{\tau}/m_H)^2 \sim 2.0 \times 10^{-4}$
and (\ref{gh temperature}).
This gives the accumulated L-asymmetry of order,
\begin{eqnarray}
&&
{\cal Y}_{BH} = c\, G_N M_{BH} \mu_L \sim  3.3 \times 10^{-10 }\, \frac{c}{16\pi}
 (\frac{\xi_4}{g} )^{4} 
\left( \frac{50\,  {\rm mg}}{M_{BH} } \right)^{17}
\,.
\end{eqnarray}
Due to the high power $\propto M_{BH}^{-17}$, there is a strong preference
of evaporating PBH mass.
There is a narrow PBH mass region around a few to several tens of
 mg that gives a right amount
of baryon to entropy ratio of order $10^{-10}$.
Generated lepton asymmetry from PBH evaporation prior to
electroweak epoch is later converted to baryon asymmetry,
although detection of the remaining lepton asymmetry is very difficult.

A problem inherent to the asymmetry generation based on
primordial black hole evaporation is that it is difficult
to reliably predict the mass distribution of potential candidates
of primordial black holes.
It is possible for a portion of primordial black holes to evaporate and
the rest to remain as dark matter, but
how much is left as dark matter is difficult to estimate.

We point out an exciting possibility.
Suppose that 
 the $N_R$ resonance temperature $T_R$
coincides with the evaporating black hole temperature $T_{BH}$, which gives
a relation,
\begin{eqnarray}
&&
M_N = 1.6 \times 10^{16} {\rm GeV}\, \frac{ 50\, {\rm mg}}{M_{BH}}
\,.
\end{eqnarray}
The preferential selection of PBH mass thus favors a particular $N_R$ mass,
thus an important microscopic parameter may be determined, though
otherwise extremely difficult to measure,
if one confirms experimentally PBH evaporation in this mass range.
Identification of PBH mass in this range may become possible
via associated strong primordial gravitational wave detection.
There may even be a possibility of detecting $e^{\pm}$  lepton asymmetry 
during $\sim 1$ year observation of evaporation,
if  PBH's of a larger mass $\sim 10^9$ gr now evaporate.

\vspace{1cm}
In summary,
the inflaton sliding, required for dynamical relaxation
towards the zero cosmological constant, gives an ideal framework
necessary to generate a finite chemical potential for lepton asymmetry.
The simplest scheme uses lepton number violating resonant scattering
that occurs in thermal medium.
The B$-$L violation works
both in thermal lepto-genesis and primordial black hole evaporation.
In either of these scenarios the favored mass scale of heavy Majorana leptons
is of order, $10^{15} \sim 10^{17}$ GeV.
Our mechanism does not require the presence of CP violating (CPV)
phases in physics beyond
the standard model, and can be realized by a minimum extension of
standard model incorporating the seesaw mechanism of small neutrino masses.
Thus, one neither needs to experimentally detect CPV phases in future
neutrino oscillation experiments,
nor CPV phases in search for electric dipole moments of various
elementary objects are necessary to identify.
It is appealing that we already know particle physics necessary
to explain matter-antimatter imbalance of our universe,
if the unified cosmology is described by multiple scalar-tensor gravity of the sort
\cite{cc relaxation}.

If the scenario of primordial black hole evaporation
is the origin of matter-antimatter imbalance,
it might be possible to determine lepton number violating heavy
Majorana masses from footprints of PBH evaporation.
With an extra assumption for simplified forms of two $3\times 3$ mass matrices,
the Majorana ${\cal M}$ and the Dirac mixing $m$, one
might be able to predict the amount of baryon asymmetry
using already known neutrino oscillation data.
These attempts are left to future works.
Nevertheless, it would  be welcome if
one experimentally finds baryon number violation along with broken B$-$L number,
since the baryo-genesis works equally well as
lepto-genesis discussed in the present work.
The discovery of proton decay certainly extends our view
on the distant future of the universe.

\section
{\bf Appendix: Thermal average of scattering rate including
the region of resonance formation}

In the energy region of $s= (p_l + p_H)^2 \geq M_N^2$
the most important is the possibility of resonance production
of heavy Majorana lepton of mass $M_N$.
We would like to estimate thermally averaged rate of scattering including
this energy region.
In order to focus on the resonance production,
we neglect much less important $u-$channel exchange contribution in
(\ref{l-violating amp}).

The squared amplitude after spin-component summation and average we consider is
\begin{eqnarray}
&&
\overline{| {\cal A}_L|^2 } = y_{HN}^4 M_N^2 \frac{ p_l\cdot \bar{p_l} }{2 E_l \bar{E_l} } 
\frac{1 }{(s - M_N^2)^2 + \gamma_N^2 M_N^2/2 }
\,.
\end{eqnarray}
Here $\gamma_N$ is the decay width of $N_R$.
In the relevant high energy region of $T \gg m_l\,, m_H$
it is legitimate to take $s\sim 2 p_l\cdot p_H$; inner product of
four-momenta  of initial lepton and Higgs boson.
Since the thermal distribution functions of bosons and fermions are
functions of their energies, we may ignore three-momenta correlation
among $\vec{p}_l \,, \vec{p}_{\bar{l}}\,, \vec{p}_H$.
This leads to an approximate formula of cross section in the free space,
\begin{eqnarray}
&&
\overline{| {\cal A}_L|^2 } \approx  \frac{ 1 }{2 } y_{HN}^4 M_N^2
\frac{1 }{(2 E_l E_H - M_N^2)^2 + \gamma_N^2 M_N^2/2 }
\,.
\end{eqnarray}

\begin{figure*}[htbp]
 \begin{center}
 \centerline{\includegraphics{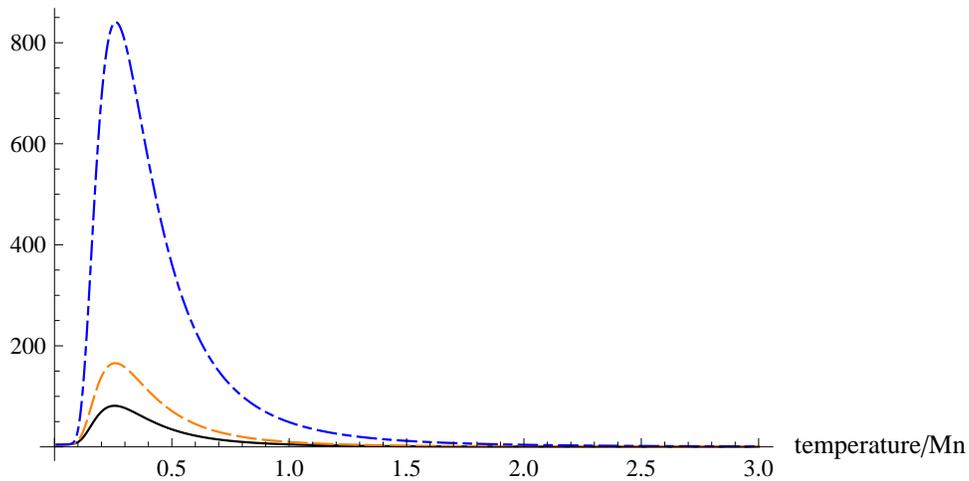}} \hspace*{\fill}
   \caption{
Lepton number violating scattering rate in thermal medium in arbitrary unit.
The function $I(x)/x^4$ of (\ref{scattering rate 0}) vs $ x = T/M_N$ for
three choices of resonance width are illustrated:
$\gamma_N/M_N = 0.1 $ in solid black, 0.05 in dashed orange,
and 0.01 in dash-dotted blue.
}
   \label {l-violating acattering rate}
 \end{center} 
\end{figure*}

The thermal average can be done taking the zero chemical potential
to a good approximation, hence one should calculate the thermally averaged
cross section $\sigma $ as
\begin{eqnarray}
&&
\sigma = \frac{1}{n_l n_H}
\int \frac{d^3 p_l d^3 p_H} {(2\pi)^6} \, \overline{| {\cal A}_L|^2 } f_B(\frac{E_H}{T}) 
f_F(\frac{E_l}{T})
\,,
\end{eqnarray}
where $f_B(x) = 1/(e^x -1)\,, f_F(y) = 1/(e^y + 1)$ and
$n_l = \zeta(3) T^3/\pi^2\,, n_H = 3 \zeta(3) T^2/ 4\pi^2$
are number densities of massless fermions and bosons.
This contains, after scaling with the temperature, an integral 
\begin{eqnarray}
&&
I (\frac{T}{M_N}) =
\int_0^{\infty}   dx \int_0^{\infty}   dy 
\frac{1 }{ (2  xy - M_N^2/T^2)^2+ \gamma_N^2 M_N^2/(2 T^4) }
\frac{x^2 y^2}{ (e^x +1)(e^y -1)}
\,,
\label {scattering rate 0}
\end{eqnarray}
times $T^{-4}$.
The averaged cross section is given by
\begin{eqnarray}
&&
\sigma \approx  \frac{ y_{HN}^4 M_N^2 }{8 \pi^4 T^4} 
I(\frac{T }{M_N }) 
\,.
\label {l-violating cross section}
\end{eqnarray}
The function $I(u)/u^4$ is illustrated in  Fig(\ref{l-violating acattering rate}).
One may  estimate the resonance temperature $T_R$, taking
the average energies for $x$ and $y$, which gives in the narrow width limit
\begin{eqnarray}
&&
T_R \approx \frac{30 \zeta(3)}{\pi^4} \sqrt{\frac{6}{7}} M_N \sim 0.34 M_N
\,.
\label {resonance temperature}
\end{eqnarray}
This gives a good approximation of what we found numerically for
the resonance position,  as in Fig(\ref{l-violating acattering rate}).
The magnitude of cross section at the resonance pole is $\propto 1/\gamma_N$,
a sensitive quantity to the decay width.
The resonance formation may be regarded as a two-step process;
the first process being Majorana $N_R$ production via inverse decay
and the second its subsequent decay.
The enhanced rate $\propto 1/\gamma_N \propto$ lifetime may be understood
by a long lifetime of Majorana particle.

The low temperature limit of $I(u)/u^4 $ is $\rightarrow 4.3$
as $u \rightarrow 0$,
while its high temperature $u \rightarrow \infty$ limit,
or $T \gg M_N$, gives
\begin{eqnarray}
&&
\frac{2.404 \ln 2}{4} \frac{1}{u^4} \sim 0.4166 \frac{1}{u^4} 
\,.
\end{eqnarray}
The thermally averaged cross sections  thus behave in these limits
\begin{eqnarray}
&&
\sigma \rightarrow 
4.3\, \frac{ y_{HN}^4 }{8 \pi^4  M_N^2} \,,
\; {\rm as}\;  \frac{T}{M_N} \rightarrow 0
\,,
\label {low energy sigma}
\\ &&
\hspace*{0.5cm}
\rightarrow
0.42\, \frac{ y_{HN}^4 M_N^2 }{8 \pi^4 T^4} \,,
\; {\rm as}\;  \frac{T}{M_N} \rightarrow \infty
\,.
\label {high energy sigma}
\end{eqnarray}
In the resonance region the integral is approximated by
\begin{eqnarray}
&&
I(u) \approx \frac{2\pi \sqrt{2} T^2}{ \gamma_N M_N} 
\int_0^{\infty} dx \int_0^{\infty} dy  \frac{(xy)^2}{ (e^x +1 ) (e^y-1 )} 
\delta (2 xy - \frac{  M_N^2 }{ T^2 })
\approx 2.4 \frac{\sqrt{2} \pi^3}{12} \frac{T^2}{\gamma_N M_N} 
\,.
\label {resonance integral} 
\end{eqnarray}
The thermally averaged cross section in the resonance region is given by
\begin{eqnarray}
&&
\sigma \approx 8.77 \, \frac{ y_{HN}^4 M_N^2 }{8 \pi^4 T^2} \frac{1}{\gamma_N M_N} \,,
\; {\rm at}\;  T = O( \frac{M_N}{3} )
\,,
\end{eqnarray}
in the narrow width limit $\gamma_N \ll M_N$.

\vspace{1cm}
 {\bf Acknowledgements}

This research was partially
 supported by Grant-in-Aid   21K03575   from the Japanese
 Ministry of Education, Culture, Sports, Science, and Technology.

\end{document}